\documentclass[%
groupedaddress,
 amsmath,amssymb,
 aps,
 prfluids,
]{revtex4-2}

\usepackage{graphicx}% Include figure files
\usepackage{dcolumn}% Align table columns on decimal point
\usepackage{bm,color}% bold math
\usepackage{hyperref}% add hypertext capabilities
\usepackage[mathlines]{lineno}% Enable numbering of text and display math
\usepackage{sansmath}

\newcommand{\tens}[1]{\mathsf{#1}}
\newcommand{\floor}[1]{\left\lfloor #1 \right\rfloor}

\newcommand{\Ca}{\mbox{\textit{Ca}}} 	% Capillary number

\newcommand{\dd}[2] { \frac{\mathrm{d} #1}{\mathrm{d} #2} }
\newcommand{\pd}[2] { \frac{\partial   #1}{\partial #2  } }
\newcommand{\pdd}[2]{ \frac{\partial^2 #1}{\partial #2^2} }

\newcommand{\Ylm}{Y^m_\ell}

\newcommand{\Uh}{\skew3\hat{\bm{v}}}

\newcommand{\urh}{\skew3\hat{v}_r}
\newcommand{\uth}{\skew3\hat{v}_\theta}
\newcommand{\uph}{\skew3\hat{v}_\varphi}
\newcommand{\ph}{\skew3\hat{p}}
\newcommand{\Pih}{\skew3\hat{\Pi}}
\newcommand{\zh}{\hat{h}}
\newcommand{\omh}{\skew3\hat{\bm{\omega}}}

\newcommand{\com}[1]{\textcolor{black}{#1}}

\newcommand*\patchAmsMathEnvironmentForLineno[1]{%
  \expandafter\let\csname old#1\expandafter\endcsname\csname #1\endcsname
  \expandafter\let\csname oldend#1\expandafter\endcsname\csname end#1\endcsname
  \renewenvironment{#1}%
     {\linenomath\csname old#1\endcsname}%
     {\csname oldend#1\endcsname\endlinenomath}}%
\newcommand*\patchBothAmsMathEnvironmentsForLineno[1]{%
  \patchAmsMathEnvironmentForLineno{#1}%
  \patchAmsMathEnvironmentForLineno{#1*}}%
\AtBeginDocument{%
\patchBothAmsMathEnvironmentsForLineno{equation}%
\patchBothAmsMathEnvironmentsForLineno{align}%
\patchBothAmsMathEnvironmentsForLineno{flalign}%
\patchBothAmsMathEnvironmentsForLineno{alignat}%
\patchBothAmsMathEnvironmentsForLineno{gather}%
\patchBothAmsMathEnvironmentsForLineno{multline}%
}

\begin{document}

\preprint{Accepted for publication in Phys. Rev. Fluids}

\title{Linear stability of ultrathin spherical coatings}
% Force line breaks with \\

\author{D. Moreno-Boza}
\email{damoreno@ing.uc3m.es}
\affiliation{Grupo de Mec\'anica de Fluidos, Departamento de Ingenier\'ia T\'ermica y de Fluidos, Universidad Carlos III de Madrid. Avda. de la Universidad 30, 28911, Legan\'es, Madrid, Spain.}

\author{A. Sevilla}
\email{alejandro.sevilla@uc3m.es}
\affiliation{Grupo de Mec\'anica de Fluidos, Departamento de Ingenier\'ia T\'ermica y de Fluidos, Universidad Carlos III de Madrid. Avda. de la Universidad 30, 28911, Legan\'es, Madrid, Spain.}
\date{\today}% It is always \today, today,

\begin{abstract}
    We unravel the linear stability properties of an otherwise stagnant ultrathin non-wetting liquid film of thickness $h_o$ coating a spherical substrate of radius $R$. The configuration is known to be unstable due to the competition of the destabilizing van der Waals (vdW) forces and the stabilizing surface tension force. The governing equations of motion written in the Stokes limit of negligible liquid inertia and an accompanying lubrication model are linearised about the zero-velocity base state and decomposed into normal modes in order to obtain the temporal dispersion relation. Discrete unstable modes are identified and tracked as a function of a capillary number $\Ca$ measuring the relative importance of surface tension to vdW forces and the film aspect ratio $\eta = R/h_o$. For small enough values of the capillary number only the first polar mode is unstable, and the corresponding maximum growth rate is shown to be a universal function of $\eta$. Lubrication theory is seen to provide a good quantitative prediction of the film stability properties for $\eta\gg 1$.
\end{abstract}

%\keywords{Suggested keywords}%Use showkeys class option if keyword display desired
\maketitle

%\tableofcontents

\section{Introduction}
Thin liquid films are found in a great number of both natural and industrial applications, and they play a central role in many manufacturing processes, physiological and medical processes, biophysics or geophysics. This has drawn the attention of relevant work over the years; for instance, falling films are relevant in coating processes~\citep{Huppert1982Nature, Kalliadasis2011}, where the ideal scenario is a stable film, spreading gravity currents are important in geological studies~\citep{Pattle1959, Huppert1982, Huppert1986,ancey2007plasticity}, pulsed-laser-heated thin films are used in patterning and plasmonic applications~\citep{Makarov2016, Hughes2017,Kondic2019}, and tear-film dynamics~\citep{Braun2012, Fuller2012, Hermans2015, Dey2019} and surfactant replacement therapy~\citep{Jensen1992, Halpern1992, Jensen1993, Halpern1993, Grotberg1994, Halpern1998, Cassidy1999} are examples where the stability of surfactant-driven films is crucial for a healthy functioning of the eye and lungs. The reader is referred to~\citep{de1985wetting, oron1997long, bonn2009wetting, craster2009dynamics, Blossey2012, Kondic2019} and references therein for a detailed review of the literature of films.

When the substrate supporting the liquid film is not planar, the classical lubrication description needs to be modified to account for the corresponding curvature distribution~\citep{schwartz1995modeling,roy2002lubrication}, which is known to significantly affect both the thin film flow and its stability~\citep{xue2021draining,ledda2022gravity}.The particular case of a thin liquid film coating a spherical surface has been studied in several contexts, ranging from the impact of a drop onto a solid spherical target~\citep{bakshi2007investigations}, to the dynamics of the canonical gravity current~\citep{takagi2010flow,balestra2018rayleigh}, including the effects sphere rotation~\citep{kang2016dynamics} and thermocapillarity~\citep{kang2017marangoni}.
 
If the thickness of the coating is sufficiently small van der Waals (vdW) forces become comparable to the surface tension force, and their influence cannot be neglected in the analysis. Indeed, for instance,~\citet{grigor1994influence} clearly showed that vdW forces profoundly affect the spectrum of unstable disturbances in electrically charged liquid shells coating a rigid spherical core. In the context of soft elastic thin films bonded to curved substrates,~\citet{boli2011} investigated the critical conditions for the occurrence of vdW-induced wrinkling using linear theory in both spherical and cylindrical configurations. The curvature of the substrate is shown to have a significant influence on the wrinkling behavior of the film. In their analysis, independence of the mode of instability with respect to azimuthal perturbations is also exhibited, which was anticipated by~\citet{colin2007morphological}.

However, to the best of our knowledge, the influence of vdW forces on the spinodal dewetting process associated to ultrathin, non-wetting liquid films coating a sphere has not been reported before. Thus, our main objective here is to present a detailed analysis of the linear stability properties of these ultrathin spherical liquid shells, using both a Stokes description and lubrication theory.

The paper is structured as follows: the general formulation of the problem is given in section~\ref{sec:formulation} followed by the linear stability analysis presented in section~\ref{sec:LSA}. Concluding remarks and future prospects are discussed in section~\ref{sec:conclusions}.

%%%%%%%%%%%% SECTION: Formulation of the problem %%%%%%%%%%%%%%

\section{Formulation of the problem}
\label{sec:formulation}
\begin{figure}[t]
    \centering
    \includegraphics[width=220pt]{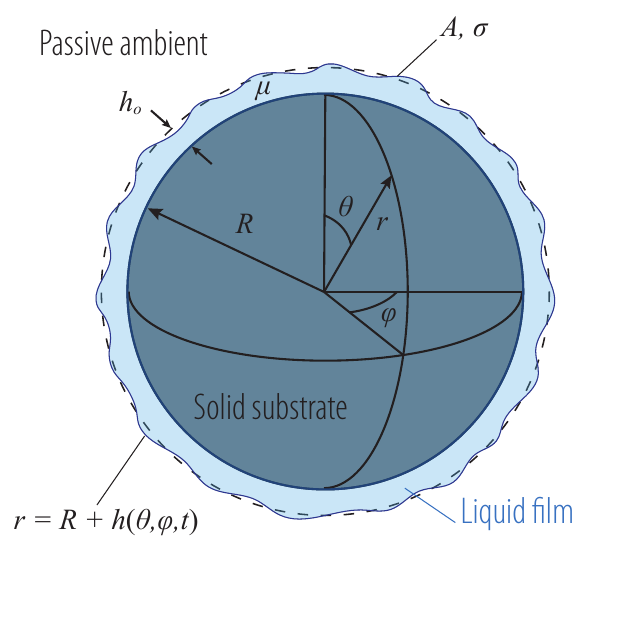}
    \caption{Schematic of the initial flow configuration, parameters, and spherical coordinates employed in the description.}
    \label{fig:sketch}
\end{figure}

A clean and initially static thin film of a Newtonian liquid of viscosity $\mu$ coating a spherical solid substrate of radius $R$ is considered. Let the instantaneous film thickness be referred to as $h$ such that its location may be defined as $r = R + h(\theta,\varphi,t)$, where spherical coordinates $(r,\theta,\varphi)$ are employed (see schematic in Fig.~\ref{fig:sketch}). The film, of initial thickness $h_o$, is surrounded by passive atmospheric air at constant pressure $p_a$. The coefficient of surface tension $\sigma$ between the air and the liquid will be taken as a constant. Finally, dispersive long-range van der Waals (vdW) forces are to be modelled through the use of a disjoining pressure $\phi(h)$ proportional to the effective Hamaker constant $A$ of the liquid-substrate pair. The initially spherical liquid-air interface may become unstable if the destabilizing vdW forces are no longer balanced by the capillary force. 

\subsection{Dimensionless equations of motion}
We make use of  
\begin{equation}
    \label{eq:scales}
     l_c = h_o, \quad v_c = A/(6 \pi \mu h_o^2), \quad t_c = 6 \pi \mu h_o^3/A, \quad p_c = A/(6 \pi h_o^3)    
\end{equation}
as relevant scales of length, velocity, time, and pressure to nondimensionalize the governing equations of motion. Since the effects of liquid inertia will be neglected under the assumption of small Laplace numbers, $\rho \sigma h_o^2/\mu \ll 1$, the motion of the liquid is effectively governed by the Stokes equations
\begin{equation}
    \label{eq:Stokes}
    \com{
    \nabla \cdot \bm{v} =  0, \quad \nabla \cdot \tens{T} = \bm{0},
    }
\end{equation}
where $\tens{T} = -p \tens{I} + \nabla \bm{v} + \nabla \bm{v}^\mathrm{T}$ is the liquid stress tensor, $p$ is the gauge pressure, and $\tens{I}$ is the identity tensor. Eq.~\eqref{eq:Stokes} needs to integrated with appropriate boundary conditions, namely, no-slip at the substrate $r = \eta$, where $\eta=R/h_o$, and the kinematic and stress balance at the liquid-air interface $r = \eta + h$,
\begin{align}
    & \bm{n} \cdot \left( \dot{\bm{x}}_s - \bm{v} \right) = 0 \label{eq:kinemBC},\\
  - & \tens{T} \cdot \bm{n} = \Ca^{-1} \, \bm{n} (\nabla_s \cdot \bm{n}) + \phi(h) \, \bm{n}, \label{eq:stressBC}
\end{align} 
where the surface has been parameterised by the position vector $\bm{x}_s$ (dot indicating time derivative) and its outward-pointing unit normal vector $\bm{n}$. The surface divergence operator is defined as $\nabla_s  = (\tens{I} - \bm{n}\bm{n})  \cdot \nabla$. It is seen that for a given initial geometrical configuration, i.e., for fixed values of $\eta$, the capillary number $\Ca = A/(6 \pi \sigma h_o^2)$ emerges as the only nondimensional parameter of the problem, measuring the relative importance of vdW and capillary forces. Conversely, one can make use of the molecular length scale $a = \sqrt{A/(6\pi\sigma)}$~\citep{gennes1985wetting,moreno2020stokes} instead of $h_o$ in order to make the problem dimensionless, \com{with} $h_o/a = \Ca^{-1/2}$ the corresponding re-scaling.

\subsection{Van der Waals potential for a liquid film coating a spherical substrate}
Equations~\eqref{eq:Stokes}--\eqref{eq:stressBC} are completed upon specification of the disjoining pressure accounting for the vdW forces. The standard expression for a planar film, $\phi_{\text{planar}} = A/(6 \pi h^3)$, would be adequate in the limit $\eta\gg 1$ of very small initial thickness compared to the substrate radius. However, it is well known that substrate curvature affects the disjoining pressure substantially~\citep{napari2003disjoining}, so that appropriate corrections are needed to describe cases where $\eta\lesssim 1$. Fortunately, the spherical shell configuration admits a closed-form description that can be determined following~\citep{israelachvili1985,boli2011} to give, in dimensional form,
\begin{equation}
    \label{eq:potentialvdWdim}
    \phi(h) = \frac{A}{6\pi h^3} \frac{8R^3}{(h+2R)^3},
\end{equation}
for a liquid coating of thickness $h$ resting on top of a spherical substrate of radius $R$. Note that $A$ may be considered either positive (dewetting) or negative (wetting) in the analysis. In the present work, we will restrict ourselves to dewetting scenarios for which $A>0$, since the wetting case is trivially stable. Upon making use of~\eqref{eq:scales}, the dimensionless version of~\eqref{eq:potentialvdWdim} is 
\begin{equation}
\label{eq:nondimpotential}
\phi(h) = \frac{8 \eta^3}{h^3 (h+2\eta)^3},
\end{equation} 
to be employed in Eq.~\eqref{eq:stressBC}. The associated small-curvature expansion is given by
\begin{equation}  
    \label{eq:phiforlargeeta}
     \phi(h) =  \frac{1}{h^3} - \frac{3}{ 2h^2 \, \eta}  +  \frac{3}{2 h \eta^2} +  O\left( \eta^{-3}  \right),
\end{equation}
in reciprocal powers of $\eta$, which effectively recovers the planar limit $\phi_{\text{planar}} = h^{-3}$ to within $O(\eta^{-1})$ relative errors. The derivative of the disjoning pressure at $h = 1$ is given by
\begin{equation}
    \label{eq:phiprimeat1}
    \phi'(1) = - \frac{48 \eta^3 (1+\eta)}{(1+2\eta)^4},
\end{equation}
which will become useful below when deriving the linearized stability equations, and which has a large-$\eta$ expansion
\begin{equation}
    \label{eq:phiprimeat1_expansion}
    \phi'(1) = - 3  +  \frac{3}{\eta} + O(\eta^{-2}).
\end{equation}

\subsection{Lubrication approximation}\label{subsec:lub}
Following a standard procedure, briefly explained in Appendix~A, the lubrication model for the evolution of the film thickness $h$ is easily deduced, yielding the evolution equation
\begin{equation}
    \label{eq:lubmodel}
    \left(\eta+h\right)^2 \sin \theta \pd{h}{t} + \pd{}{\theta} \left(  G(h,\eta) \, \sin\theta \, \pd{p}{\theta} \right) + \frac{1}{\sin\theta} \pd{}{\varphi} \left( G(h,\eta) \pd{p}{\varphi} \right) = 0,
\end{equation} 
with mobility function $G$ and static pressure $p$ computed through
\begin{equation}
    \label{eq:lubdefpress}
    G(h,\eta) = \frac{h^3(h + 2 \eta)}{6 (h-\eta)}, \quad p = \frac{\mathcal{C}}{\Ca} + \phi, 
\end{equation} 
where the curvature $\mathcal{C}$, whose exact expression is provided as Eq.~\eqref{eq:curvature_complete} in Appendix~\ref{app:A}.
The lubrication model given by Eq.~\eqref{eq:lubmodel} was also deduced by~\citet{kang2016dynamics} (see also~\citep{kang2017marangoni}). Note that, although Eq.~\eqref{eq:lubmodel} assumes slender flow, the effect of substrate curvature is retained not only through the metric factor $\sin\theta$, but also through the functions of $\eta$ appearing in the first term and in the mobility function $G(h,\eta)$. The classical thin-film approximation~\citep{kang2016dynamics,balestra2018rayleigh} is obtained by expanding Eq.~\eqref{eq:lubmodel} in powers of $\eta^{-1}$ for $\eta\gg 1$ to yield, at leading order,
\begin{equation}
    \label{eq:thinfilmmodel}
    \eta^2 \sin \theta \pd{h}{t} - \pd{}{\theta} \left(  \sin\theta \, \frac{h^3}{3} \pd{p}{\theta} \right) - \frac{1}{\sin\theta} \pd{}{\varphi} \left( \frac{h^3}{3} \pd{p}{\varphi} \right) = 0,
\end{equation} 
with pressure now given by $p = \mathcal{C}/\Ca + h^{-3}$, and curvature $\mathcal{C}$ linearized around the base state according to
\begin{equation}\label{eq:curv}
\mathcal{C} =  \frac{2}{(1+\eta)} - \frac{1}{(1+\eta)^2} \left[ \frac{1}{\sin^2{\theta}} \frac{\partial^2 h}{\partial \varphi^2} +  \frac{1}{\sin \theta}  \pd{}{\theta} \left(\sin\theta \pd{h}{\theta} \right)  + 2 h \right]. 
\end{equation} 
\com{Note that the prefactor in Eq.~\eqref{eq:thinfilmmodel} calls for the temporal rescaling $t \to t/\eta^2$, which would render it parameter-free. However, curvature effects must be retained in the definition of the capillary pressure through the curvature in Eq.~\eqref{eq:curv} so as to produce the cut-off of temporal instability.
}
\section{Linear stability}\label{sec:LSA}

\subsection{Cut-off polar wavenumber}
As shown in Appendix~\ref{app:B}, a Plateau-like free energy argument allows the derivation of the neutral stability condition, which reads
\begin{equation}
\label{eq:cutoff}    
\ell (\ell + 1)  =  2  - \phi'(1) \Ca (\eta+1)^2,
\end{equation} 
where $\ell=0,1,2,\ldots$ is the polar wavenumber of the spherical harmonic and $\phi'(1)=- 48 \eta^3 (1+\eta)/(1+2\eta)^4<0$ in the non-wetting case considered herein. The cut-off polar wavenumber $\ell_c$ is then given by 
\begin{equation}
\label{eq:cutoff_solution}
    \ell_c = \floor{ \frac{1}{2} \left(\sqrt{ 9 - 4  \Ca \, \phi'(1) (\eta +1)^2 }-1 \right) },
\end{equation} 
where $\floor{\cdot}$ represents integer part, revealing that perturbations with $0 < \ell < \ell_c$ and $|m| \leq \ell$, are linearly unstable. Note that the spherically symmetric mode $\ell=0$ is always neutrally stable, the mode $\ell=1$ is unconditionally unstable, and the value of $\ell_c\geq 2$ increases with $\Ca$ and $\eta$. In the limit of very thin films, $\eta\gg 1$, it proves convenient to define $k=\ell/\eta$ as the equivalent wavenumber of a planar film. Indeed, a large-$\eta$ expansion of~\eqref{eq:cutoff_solution} provides $k_c = \ell_c/\eta = \sqrt{3 \Ca} + O(\eta^{-1})$, effectively recovering the classical cutoff of a planar film~\citep{scheludko1962certaines,vrij1966possible}.

\subsection{The dispersion relation}
The governing equations~\eqref{eq:Stokes} and the boundary conditions~\eqref{eq:stressBC} are decomposed into normal modes by introducing the usual ansatz $\bm{q}(\bm{x},t) = \bm{Q}(\bm{x}) + \epsilon \hat{\bm{q}}(\bm{x}) \exp{( \omega t)}$ where $\bm{q} = (v_r,v_\theta,v_\varphi,p,h)$ is the vector of state variables, $\omega$ is the growth rate, $0 < \epsilon \ll 1$ is a small amplitude and the steady-state solution is $\bm{Q} = (0,0,0,p_o,1)$. Note that the dependence of the eigenfunctions on the spatial coordinates has not been yet anticipated. Upon substitution of the ansatz in Eq.~\eqref{eq:Stokes} one finds
\begin{equation}
    \label{eq:stokeslin}
    \nabla \cdot \Uh = 0, \quad \bm{0} = -\nabla \ph + \nabla^2  \Uh, 
\end{equation} 
valid in the spherical shell $\eta < r < \eta+1$, $0<\theta<\pi$, $0<\varphi < 2\pi$. Following the procedure explained in Appendix~\ref{app:C}, a closed-form solution to the associated dispersion relation for the growth rate $\omega$ is found as a rational function of $\ell$, $\eta$ and $\Ca$, given by
\begin{equation}
    \label{eq:omspherical}
    \omega = \frac{\ell (\ell+1) (2 \ell+1)}{2 \Ca (\eta + 1) } 
    \, \times \,
    \frac{  B_1   \left[ \ell (\ell+1) - 2  +  \phi'(1) \Ca \, (1+\eta)^2 \right] }
    { B_2 + \left( B_3 + B_4    + B_5 + B_6 \right) \eta ^{2 \ell} (\eta + 1)^{2 \ell}  + B_7} 
\end{equation}
where the polynomials $B_i,\,i=1\ldots 7$, provided in Appendix~\ref{app:D}, have been introduced for convenience.

It may be recognised that $\omega > 0$ for $\Ca > 0$ and a range of discrete positive values of $0 < \ell  \leq \ell_c$, where the cut-off wavenumber $\ell_c$, defined as the largest value of $\ell$ for which $\omega>0$, is given by Eq.~\eqref{eq:cutoff_solution}. The cut-off value can be found by inspection of the linearized curvature term, which appears in Eq.~\eqref{eq:omspherical} multiplying the polynomial $B_1$, providing the same cut-off as that found above in~\eqref{eq:cutoff}. Since the azimuthal wavenumber does not appear explicitly in the dispersion relation, the flow is unstable for all values of $|m| \leq \ell$, provided that $\ell \leq \ell_c$. Note that a similar result is obtained, for instance, by~\cite{colin2007morphological} in the context of stressed composite spherical shells and by~\cite{boli2011} for the vdW-induced instability of thin elastic spherical shells.

The limit $\eta \gg 1$ of a liquid film coating a planar substrate may be effectively recovered by noticing that as $\eta$ grows, the number of unstable polar modes accommodated must grow accordingly. According to the curvature term in~\eqref{eq:eigbcstress}, $\ell \sim \eta$ as $\eta \to \infty$ in order to produce a finite wavenumber $k = \ell/\eta$, which plays the role of the streamwise wavenumber defined in~\cite{Jain1976}. Indeed, the planar growth rate
\begin{equation}
    \label{eq:omplanar}
    \omega_{\text{planar}} = \frac{(k^2 - 3 \Ca)(k - \cosh{k} \sinh{k})}{ 2 k \Ca (k^2 + \cosh^2{k} )} 
\end{equation}
(see for instance \cite{Ruckenstein1974}) may be obtained by taking the double limit $\eta, \ell \to \infty$ of~\eqref{eq:omspherical} for finite $\ell/\eta$ and retaining only the leading-order term in~\eqref{eq:phiforlargeeta}. The classical result for the cut-off wavenumber $k_c = \sqrt{3 Ca}$ is recovered, originally deduced by~\citet{scheludko1962certaines} and~\citet{vrij1966possible} for planar films. The required algebra for such simplification, which is not trivial, has been carried out with the help of~\citep{Mathematica}.

In the case of the lubrication models~\eqref{eq:lubmodel} and~\eqref{eq:thinfilmmodel}, the associated dispersion relations are given by
\begin{equation}
    \label{eq:lubdisprel}
    \omega_{\text{LUB}_1} = \frac{(2 \eta +1) \, \ell (\ell+1) \left[ 2 - \phi'(1) \Ca \, (\eta+1)^2 -  \ell (\ell+1)  \right]}{6 \Ca \, (\eta -1) \,  (\eta +1)^4},
\end{equation}
and
\begin{equation}
    \label{eq:reynoldsDR}
    \omega_{\text{LUB}_2} = \frac{\ell (\ell+1) \left[ 2 + 3 {Ca} (\eta +1)^2 - \ell(\ell+1)  \right]}{3 {Ca} \,  \eta ^2 (\eta +1)^2},
\end{equation}
respectively, where, in the latter case, it has been taken into account that $\phi'(1)=-3$ in the thin-film limit $\eta\gg 1$. 

An unexpected feature of the lubrication model with substrate curvature effects is the existence of a singularity at $\eta=1$, manifested by the factor $\eta-1$ appearing in the denominator of Eq.~\eqref{eq:lubdisprel}. This failure can be traced back to the solution of the velocity field in spherical coordinates, given by Eqs.~\eqref{eq:vtheta_lub}-\eqref{eq:vphi_lub}, where the factor $h-\eta$ appears in the denominator as a consequence of the imposition of the leading-order expression for the stress-free boundary condition at the liquid-air interface. This singularity represents a major limitation for the practical use of the leading-order lubrication model for relatively thick films which, in particular, predicts an infinite disturbance growth rate for $\eta=1$, and a negative growth rate for $\eta < 1$. As shown in Appendix~\ref{app:E}, a simple regularization of the singularity can be achieved simply by retaining the radial velocity of the interface in the stress-free boundary condition. In contrast, such a singularity is absent in the thin-film lubrication model~\eqref{eq:reynoldsDR} although, as shown below, and not unexpectedly, the latter model also fails catastrophically in the prediction of the stability properties of the film for $\eta\lesssim 1$.

\begin{figure}
    \centering
    \includegraphics[width=420pt]{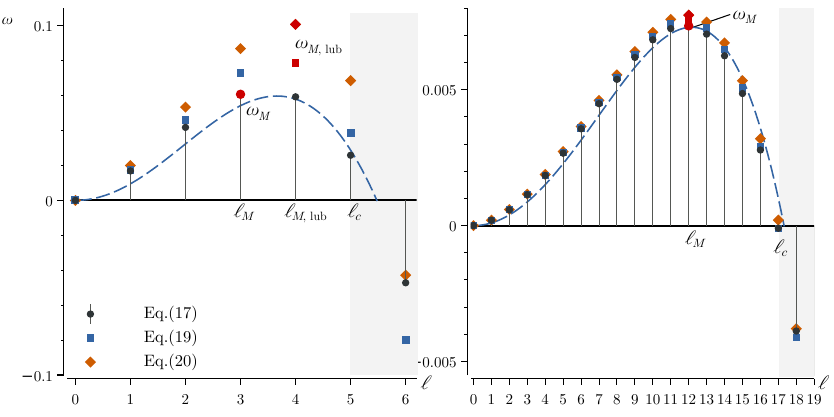}
    \caption{(a) Temporal growth rate for (a) $\Ca = 0.1, \eta=10$ ($\ell_c=5$) and (b) $\Ca = 0.01$, $\eta = 100$ ($\ell_c=17$). Symbols represent the results obtained using the Stokes dispersion relation~\eqref{eq:omspherical} (circles), the lubrication model with curvature effects~\eqref{eq:lubdisprel} (squares) and the thin-film lubrication model~\eqref{eq:reynoldsDR} (diamonds). The planar limit given by~\eqref{eq:omplanar} is also represented for continuous values of $\eta k$ and $0 < k < \sqrt{3\ Ca}$ (dashed line).}
    \label{fig:fig2}
\end{figure}

Figure~\ref{fig:fig2} shows the temporal growth rate $\omega(\ell)$ for two different cases, namely (a) $Ca=0.1,\,\eta=10$, for which $\ell_c=5$, and (b) $Ca=0.01,\,\eta=100$, for which $\ell_c=17$. Four results are displayed in Fig.~\ref{fig:fig2}, corresponding to the Stokes dispersion relation~\eqref{eq:omspherical} (circles), the lubrication model with substrate curvature effects~\eqref{eq:lubdisprel} (squares), the lubrication model in the thin-film limit~\eqref{eq:reynoldsDR} (diamonds) and the planar limit~\eqref{eq:omplanar} (dashed line). The maximum growth rate is highlighted in red. Both lubrication models overpredict the temporal growth rate considerably, being the model without curvature effects the worst performing one. Moreover, in the case of the relatively thicker film with $\eta=10$, both lubrication models fail to account for the optimal polar wavenumber, since they predict $\ell_M=4$, while the Stokes dispersion relation yields $\ell_M=3$. Interestingly, the equivalent planar dispersion relation~\eqref{eq:omplanar} is able to reproduce the exact growth rate fairly well even in the case with $\eta=10$, since it does not rely on the slenderness of the self-induced flow, but only on the weak curvature approximation. In the limit of very thin films, exemplified by the case with $\eta=100$, the three models are seen to correctly capture the exact result with small relative errors within the whole range of unstable wavenumbers, including the prediction of $\ell_M=12$.

\begin{figure}
    \centering
    \includegraphics[width=420pt]{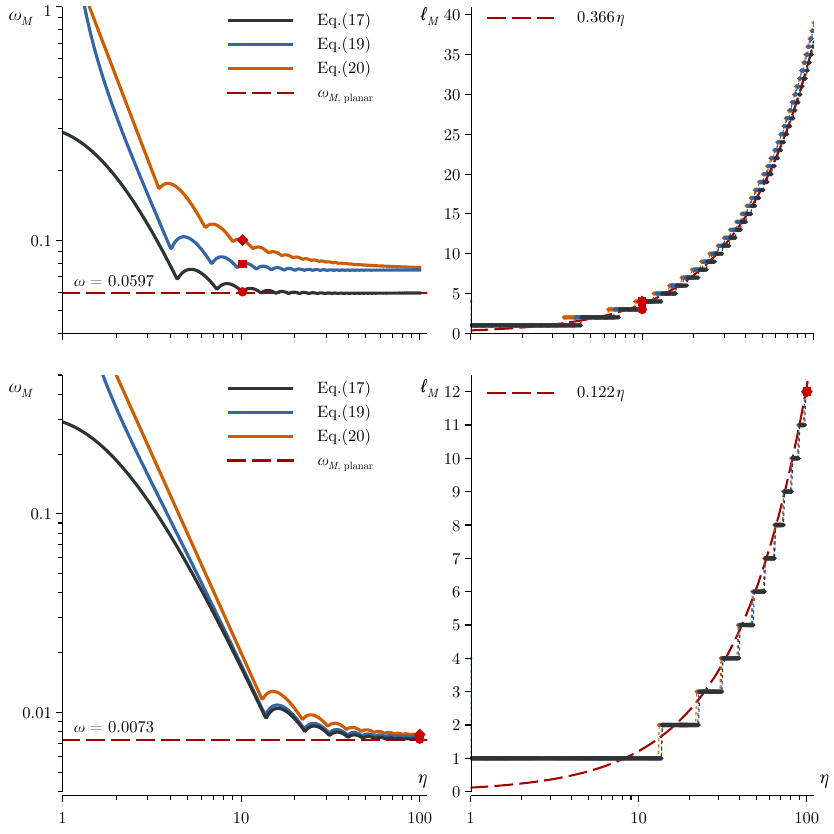}
    \caption{Maximum growth rate $\omega_M$ (left panels) and corresponding optimal wavenumber $\ell_M$ (right panels) as functions of the aspect ratio $\eta$ for $Ca = 0.1$ (upper panels) and $Ca = 0.01$ (lower panels), as predicted by the Stokes dispersion relation~\eqref{eq:omspherical} (black solid lines), the lubrication model with curvature effects~\eqref{eq:lubdisprel} (blue solid lines), the thin-film lubrication model~\eqref{eq:reynoldsDR} (orange solid lines), and the planar dispersion relation~\eqref{eq:omplanar} (dashed lines). Red symbols indicate the values of $\omega_M$ and $\ell_M$ corresponding to Fig.~\ref{fig:fig2}.}
    \label{fig:fig3}
\end{figure}

The main result of the present paper is summarized in Fig.~\ref{fig:fig3}, which shows the maximum growth rate $\omega_M$ (left panels) and the corresponding optimal wavenumber $\ell_M$ (right panels) as functions of $\eta$, for $\Ca=0.1$ (upper panels) and $\Ca=0.01$ (lower panels). The exact result, obtained from the Stokes dispersion relation (black solid lines), clearly shows that the function $\omega_M(\eta)$ has discontinuous derivatives at discrete values of $\eta=\eta_i(\Ca),\,i=1,2,\ldots$, associated with the monotonic increase of the dominant wavenumber $\ell_M=i$ as $\eta$ increases. For $\Ca=0.1$, the Stokes dispersion relation provides $\eta_i=(4.38,7.31,10.05,12.89,\ldots)$, corresponding to changes of the dominant wavenumber from $\ell_M=1$ for $\eta<4.38$, to $\ell_M=2$ for $4.38<\eta<7.31$, and so on. These results are to be compared with the lubrication results, namely $\eta_{i,\text{LUB}_1}=(4.03,6.82,9.51,12.03,\ldots)$ and $\eta_{i,\text{LUB}_2}=(3.46,6.27,8.99,11.54,14.20,\ldots)$, which clearly show the better relative performance of the lubrication model with account taken for the substrate curvature. In the case $\Ca=0.01$, we obtain $\eta_i=(13.62,22.74,31.26,39.55,\ldots)$ using the Stokes dispersion relation, while the lubrication models provide $\eta_{i,\text{LUB}_1}=(13.62,22.42,30.83,39.01,\ldots)$ and $\eta_{i,\text{LUB}_2}=(13.07,21.81,30.41,38.470,\ldots)$ revealing that, as expected, the performance of both lubrication models improves as $\Ca$ decreases and $\eta$ increases. Another important conclusion drawn from Fig.~\ref{fig:fig3} is that lubrication theory must be used with great care in spherical geometries if quantitative precision is desired. Indeed, for $\Ca=0.1$, the results of the lubrication models are seen to substantially overpredict the maximum growth rate over the whole range of values of $\eta$, especially for $\eta\lesssim 10$. For $\eta\to\infty$, both lubrication models reach an asymptotic value that is larger than the exact value provided by the spherical and planar Stokes dispersion relations. For a smaller value of $\Ca=0.01$ the characteristic streamwise length of the most unstable mode is larger, and thus the lubrication prediction improves with respect to the case with $\Ca=0.1$, especially for $\eta\gg 1$. Nevertheless, the quantitative agreement deteriorates with decreasing values of $\eta$, and is completely lost for $\eta\lesssim 3$. Finally, it is interesting to note that the equivalent planar dispersion relation~\eqref{eq:omplanar} correctly reproduces the exact result for $\eta\gg 1$ and arbitrary values of $\Ca$, since it does not rely on the slenderness of the flow.

\begin{figure}
    \centering
    \includegraphics[width=320pt]{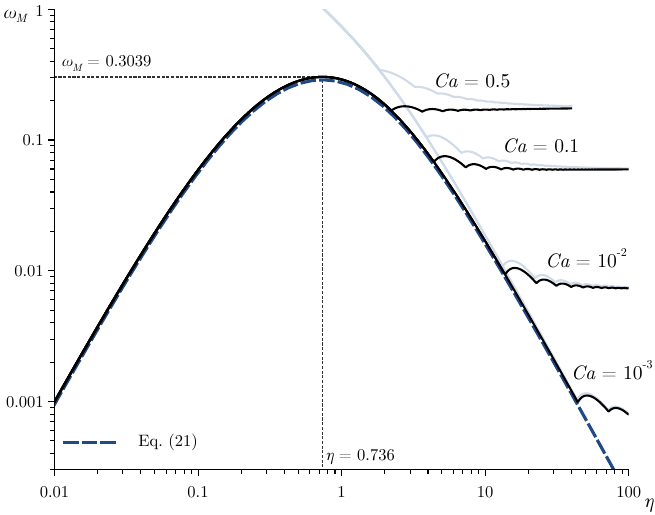}
    \caption{Optimal growth rate for the Stokes dispersion relation for several values of $Ca$ using the complete expression for $\phi'(1)$ (solid black lines) and $\phi'(1) \simeq -3$ (solid light grey lines). The envelope function given by Eq.~\eqref{eq:function_of_eta_nested} is represented for visual guidance (dashed blue line).}
    \label{fig:fig4}
\end{figure}

Since the polar mode $\ell=1$ is the dominant one for a substantial range of film aspect ratios, it proves convenient to particularize the Stokes dispersion relation~\eqref{eq:omspherical} to $\ell=1$, yielding the universal function
\begin{equation}
    \label{eq:function_of_eta_nested}
    \omega_{\ell=1}(\eta) = \frac{16 \eta ^2 (\eta +1)^2 \left[5 \eta \, (\eta +1)+2\right]}{(2 \eta +1)^3 
    \left[5 \eta \,  \left(\eta  \, (\eta \,   (\eta \,  (\eta +3)+6)+6)+3 \right)+3\right]}
    %= 
    %\underbrace{
    %\com{\frac{2}{\eta ^2}-\frac{3}{\eta ^3}+O(\eta^{-4})}
    %}_{\text{Interestingly enough}}
\end{equation}
represented with a dashed line in Fig.~\ref{fig:fig4}, which, interestingly enough, is independent of $\Ca$. The function $\omega_{\ell=1}$ displays a maximum value $\omega=0.3039$ at $\eta=0.736$, and is seen to correctly capture the linear stability properties of the film for $\eta\lesssim 2$ and any realistic value of $\Ca$. Indeed, note that $\Ca=1$ corresponds to a film whose initial thickness $h_o$ is equal to the molecular length scale $a$. The smaller the value of $\Ca$, the larger the maximum value of $\eta=\eta_{\text{max}}$ for which Eq.~\eqref{eq:function_of_eta_nested} correctly represents the linear stability properties. For instance, $\eta_{\text{max}}=(4.38,13.62,44.2)$ for $\Ca=(0.1,0.01,0.001)$, corresponding to the values of $\eta$ at which the dominant $\ell$-mode jumps by one unit. For $\eta\ll 1$, Eq.~\eqref{eq:function_of_eta_nested} has the expansion
$\omega_{\ell=1}=32 \eta ^2/3-208 \eta ^3/3+O(\eta^4)$, revealing a strong stabilization effect for very thick films. 

To assess the role of the vdW potential, Fig.~\ref{fig:fig4} also includes the result of a calculation where the exact expression for $\phi'(1)$, given by Eq.~\eqref{eq:phiprimeat1}, is substituted by its thin-film approximation of -3, corresponding to the first term of the expansion~\eqref{eq:phiprimeat1_expansion}. The result, represented with a gray solid line, reveals that a proper description of thin films with aspect ratios $\eta\lesssim 10$ requires the use of the exact vdW potential, which cannot be substituted by its thin-film approximation if quantitative precision is required.

\section{Conclusions}\label{sec:conclusions}
We have presented a detailed analysis of the linear stability properties of films of a Newtonian liquid coating the external surface of a solid spherical substrate under conditions in which the inertia of the liquid is negligible and van der Waals (vdW) forces are comparable to surface tension forces. The linearized Stokes equations have been solved analytically to yield an exact dispersion relation describing the set of discrete temporal modes of the liquid film in terms of spherical harmonics parameterized by the polar wavenumber $\ell$. Two nondimensional parameters govern the flow, namely a capillary number $\Ca$, which compares the relative strength of the vdW and surface tension forces, and the ratio of the sphere radius to the initial film thickness, $\eta$. The first polar mode, $\ell=1$, has been shown to be unconditionally unstable, while higher-order modes with $\ell\geq 2$ are destabilized when either $\Ca$ or $\eta$ become large enough. The dominant mode $\ell_M$ has been determined by solving the dispersion relation in wide ranges of the $(\eta,\Ca)$ parameter plane. For a given value of $\Ca$, the value of $\ell_M$ has been shown to increase in unit steps at a discrete set of values of $\eta$. In the limit of very thin films, $\eta\gg 1$, the stability properties converge to those of a planar film with an equivalent continuous streamwise wavenumber $k=\ell/\eta$ defined in the double limit $\eta,\ell\to\infty$. In the limit of thick films, $\eta\lesssim 1$, the dominant mode is always $\ell_M=1$ for realistic values of $\Ca$, and the maximum growth rate becomes a universal function of $\eta$ displaying a maximum of $0.3039$ at $\eta=0.736$. We have also shown that the use of the exact expression of the spherical vdW potential is mandatory for films with aspect ratio $\eta\lesssim 10$, which cannot be described using the standard simplification routinely used in the literature.

The performance of lubrication theory, which is routinely employed to describe the dynamics of the type of liquid films studied herein, has also been assessed using two different approximations: a first model that retains the curvature of the spherical substrate, and a second one valid in the limit $\eta\gg 1$ of thin films relative to the substrate. The latter model, which is the one used in previous works, has been shown to provide a reasonably good quantitative agreement with the Stokes equations, but only for small enough values of $\Ca$ and large enough values of $\eta$. The first model improves the agreement with the full equations of motion for $\eta>1$, although it displays an unexpected singularity at $\eta=1$ related with the structure of the velocity field in the lubrication approximation when the stress-free boundary condition is imposed at the liquid-air interface. A regularization of this singularity is possible if the radial velocity is retained in the shear-free boundary condition, although the relative errors of the regularized model remain large when applied to relatively thick films with $\eta\lesssim 1$, which can only be properly described with use of the full equations of motion. The failure of the lubrication models to account for the linear stability properties of relatively thick films rises doubts about their validity for the description of the nonlinear dynamics of the film, which will be reported in a separate contribution.

%-----ACKNOWLEDGEMENTS------%

\begin{acknowledgments}
The authors thank Dr. A. Mart\'inez-Calvo for insightful discussions. This research was funded by the Spanish MCIU-Agencia Estatal de Investigaci\'on through project PID2020-115655GB-C22, partly financed through FEDER European funds.
\end{acknowledgments}

%%%%%%%%%%%%%%%%%%%%%%%%%%%%%%%%%%%%%%%%%%%%%%%%%%%%%%

\begin{appendix}

\section{Derivation of the lubrication models}\label{app:A}

We follow the steps of~\citet{kang2016dynamics} in order to derive a nonlinear equation for the evolution of the film height $h(\theta,\varphi,t)$. We begin by writing the kinematic condition in the liquid-air interface $r=\eta + h$, 
\begin{equation}
    \label{eq:kineminterface}
    \pd{h}{t} = v_r - \frac{v_\theta}{r}\pd{h}{\theta} - \frac{v_\varphi}{r \sin\theta} \pd{h}{\varphi}, 
\end{equation} which combined with the continuity equation after integrating across the liquid layer results in
\begin{equation}
    \label{eq:continuityaveraged}
    (\eta + h)^2 \sin\theta \pd{h}{t} + \pd{}{\theta} \left( \sin\theta \int^{\eta+h}_{\eta} r v_\theta \, \mathrm{d}r \right) + \pd{}{\varphi} \int^{\eta+h}_{\eta} r v_\varphi \, \mathrm{d}r = 0.
\end{equation} The radial-component of the momentum equation yields $\partial_r p = 0$ to leading order, therefore $p = p(\theta,\varphi,t)$. Direct integration of the in-plane components of the momentum equation across the liquid layer is used to determine the polar and azimuthal components of the velocity, namely
\begin{align}
    \frac{1}{r} \pd{p}{\theta} & = \frac{1}{r^2} \pd{}{r} \left( r^2 \pd{v_\theta}{r}  \right), \label{eq:lub_thetamom}\\
    \frac{1}{r \sin\theta} \pd{p}{\varphi} & = \frac{1}{r^2} \pd{}{r} \left( r^2 \pd{v_\varphi}{r}  \right), \label{eq:lub_phimom}
\end{align} for which
\begin{align}
    & v_\theta  = \frac{\partial p/\partial \theta}{2 r}  \, \frac{r-\eta}{h-\eta}  \left[ h(r+\eta) - r\eta + \eta^2 \right], \label{eq:vtheta_lub} \\
    & v_\varphi = \frac{\partial p/\partial \varphi}{2 r \sin\theta} \,    \frac{r-\eta}{h-\eta} \left[ h(r+\eta) - r\eta + \eta^2 \right], \label{eq:vphi_lub}
\end{align} where the no-slip boundary condition at $r = \eta$, $v_\theta = v_\varphi = 0$, and the leading-order zero-tangential-stress boundary condition at $r = \eta + h$, $\partial_r \left(v_\theta/r \right) = \partial_r \left(v_\varphi/r \right)  = 0$, have been imposed. Inserting the expressions for the velocities into~\eqref{eq:continuityaveraged} and integrating yields 
\begin{equation}
    \label{eq:lubmodel_app}
    \left(\eta+h\right)^2 \sin \theta \pd{h}{t} + \pd{}{\theta} \left(  G(h,\eta) \, \sin\theta \, \pd{p}{\theta} \right) + \frac{1}{\sin\theta} \pd{}{\varphi} \left( G(h,\eta) \pd{p}{\varphi} \right) = 0,
\end{equation} where the auxiliary mobility function $G$ is defined in~\eqref{eq:lubdefpress}. This evolution equation for the film height must be complemented with the normal-stress condition at leading order $p = \mathcal{C} + \phi$, where the complete curvature $\mathcal{C}$ is given by
{\small
\begin{equation}\label{eq:curvature_complete}
\mathcal{C} = 
    \frac{h_\theta \left\{ 3 h_\theta r_s - [ h_\theta^2 + r_s^2]  \cot\theta \right\}
    - r_s^2 h_{\theta\theta}     
    - \left\{ h_{\varphi\varphi} \left[h_\theta^2+r_s^2\right]
             +h_\varphi^2 \left[ h_{\theta\theta} + 2 h_\theta  \cot\theta  - 3 r_s\right]
             + 2 h_\varphi h_\theta h_{\theta\varphi} 
        \right\}  \csc^2 \theta     
    + 2 r_s^3}
    {r_s \left[h_\theta^2+\csc^2\theta h_\varphi^2+r_s^2\right]^{3/2}}
\end{equation}
}
with $r_s = \eta + h$.

\section{Calculation of the critical conditions}\label{app:B}
Following the Plateau-like argument given in~\cite{scheludko1962certaines,vrij1966possible, vrij1968rupture}, in which undulations of the free surface with negative extra Gibbs energy tend to grow in time, let us consider the dimensionless excess Gibbs energy $\Delta G$ for our particular configuration. To that end, let $\Delta G = \Delta G_s + \Delta G_\textrm{molec}$, where $\Delta G_s = \Ca^{-1} \Delta S$ is the Gibbs energy associated to the extra surface area $\Delta S$ due to the perturbation and $\Delta G_\textrm{molec}$ is the extra interaction energy due to the molecular interactions, which take the form
\begin{align}
    & \Delta S = \int^{2\pi}_0 \!\!\! \int^\pi_0  \left| \bm{X}_\theta \wedge \bm{X}_\varphi \right| \, \mathrm{d}\theta \mathrm{d}\varphi - 4 \pi (\eta+1)^2, 
    \label{eq:excessS}\\
    & \Delta G_\textrm{molec} = \int^{2\pi}_0 \!\!\! \int^\pi_0  \left[ \phi(1) \Delta h + \phi'(1)(\Delta h)^2/2 + \ldots \right] (\eta+1)^2 \sin\theta \, \mathrm{d}\theta \mathrm{d}\varphi, \label{eq:excessG}
\end{align} 
where $\bm{X} = r(\theta,\varphi) ( \sin\theta \cos\varphi, \sin\theta \sin\varphi , \cos\theta)$ is the Cartesian parameterization of the corrugated surface $ r = r(\theta,\varphi) = \eta + h(\theta,\varphi) = \eta + 1 + \epsilon Y^m_\ell (\theta,\varphi)$, $\Ylm$ are the spherical harmonics of degree $\ell \geq 0$ and order $-\ell \leq m \leq \ell$ (integer numbers), $\Delta h = h-1 = \epsilon Y^m_\ell(\theta,\varphi)$, and $V(h) = -1/(2h^2)$ is the energy of interaction per unit area among the molecules in the film, assumed to be essentially determined by van der Waals forces, such that $V'(h) = \phi(h)$. Note that $V''(1) = \phi'(1)$ which for $\eta \gg 1$ gives $\phi'(1) = -3 + O(1/\eta)$, according to Eq.~\eqref{eq:phiforlargeeta}. Eqs.~\eqref{eq:excessS}--\eqref{eq:excessG} must be complemented with the incompressibility condition
\begin{equation}
    \label{eq:excessV}
    \Delta \mathcal{V} =  \int^{2\pi}_0 \!\!\! \int^\pi_0  \!\!\! \int^{\eta+1+\epsilon Y_\ell^m}_\eta \!\!\!\! r^2 \sin\theta \,  \mathrm{d}r \mathrm{d}\theta \mathrm{d}\varphi - \frac{4}{3}\pi \left[ (\eta + 1)^3 - \eta^3 \right] = 0,
\end{equation} 
where $\mathcal{V}$ is the volume of the liquid shell, which yields the following condition between the first-order and the higher-order contributions, namely,
\begin{equation}
    \label{eq:excess_incompressibilitycondition}
    \epsilon (\eta+1) \int^{2\pi}_0 \!\!\! \int^\pi_0 Y_\ell^m \sin\theta \, \mathrm{d}\theta \mathrm{d}{\varphi} = - \epsilon^2   \int^{2\pi}_0 \!\!\! \int^\pi_0 (Y_\ell^m)^2 \sin\theta \, \mathrm{d}\theta \mathrm{d}{\varphi}  + O(\epsilon^3)
\end{equation} 
Upon introducing~\eqref{eq:excess_incompressibilitycondition} into~\eqref{eq:excessS}, the excess surface turns out to be
\begin{align}
    \Delta S & =  \epsilon^2  \!  \int^{2\pi}_0 \!\!\! \int^\pi_0   \left[ - (Y_\ell^m)^2 + \frac{1}{2}\left( \frac{1}{\sin\theta} \pd{Y_\ell^m}{\varphi} \right)^{\!\!2} \! + 
    \frac{1}{2}\left( \pd{Y_\ell^m}{\theta} \right)^{\!\!2} \right] \sin\theta \, \mathrm{d}\theta \mathrm{d}\varphi + O(\epsilon^3) \nonumber   \\ 
            & =  
             \frac{\epsilon^2}{2} \left[  \ell (\ell+1) - 2\right] + O(\epsilon^3),
\end{align} 
where the integrals have been performed leveraging the orthogonality properties of the spherical harmonics and the results found in~\citet{barrera1985vector}, namely, 
\begin{equation}
    \int_\Omega (Y_\ell^m)^2\,\mathrm{d}\Omega = 1, \quad \text{and} \quad 
    \int_\Omega  |r \nabla Y^m_\ell|^2 \, \mathrm{d}\Omega = \ell (\ell + 1),
\end{equation} 
where $\mathrm{d} \Omega = \sin\theta \mathrm{d}\theta \mathrm{d}\varphi$ is the spherical solid angle element. The $O(\epsilon^2)$-contribution to the Gibbs energy is
\begin{equation}
    \Delta G_\textrm{molec} =  \frac{\phi'(1)}{2} (\eta+1)^2 \epsilon^2 \int^{2\pi}_0 \!\!\! \int^\pi_0 (Y_\ell^m)^2 \sin\theta \, \mathrm{d}\theta \mathrm{d}\varphi =   \frac{\phi'(1)}{2} (\eta+1)^2 \epsilon^2.
\end{equation} Finally, any perturbation is allowed to grow exponentially if the total excess energy $\Delta G$ becomes negative. The cutoff is thus found when $\Delta G = 0$ or, equivalently, to second order,
\begin{equation}
    0 = \Ca^{-1} \Delta S + \Delta G_\textrm{molec} = \frac{\epsilon^2}{2 \Ca } \left[  \ell (\ell+1) - 2\right] +  \frac{\phi'(1)}{2} (\eta+1)^2 \epsilon^2 = 0,
\end{equation} 
for which the cutoff wavenumber $\ell_c$ is readily obtained as the positive solution (recall that $\phi'(1)<0$) to the equation 
\begin{equation}
\label{eq:cutoff_app}    
\ell_c (\ell_c + 1)  - 2  + \phi'(1) \Ca (\eta+1)^2 = 0,
\end{equation} 
which is given by 
\begin{equation}
\label{eq:cutoff_solution_app}
    \ell_c = \floor{ \frac{1}{2} \left(\sqrt{ 9 - 4  \Ca \, \phi'(1) (\eta +1)^2 }-1 \right) },
\end{equation} 
where $\floor{\cdot}$ represents integer part, revealing that perturbations with $0 < \ell < \ell_c$ and $|m| \leq \ell$, are linearly unstable.

\section{Solution of the linearised Stokes equations}\label{app:C}
\label{app:poloidaltoroidal}

The perturbed apparent pressure $\Pih = \ph + \phi'(1) \zh $, where $\phi'(1) = \mathrm{d}\phi/\mathrm{d}h|_{h=1}$ is specified in the main text, satisfies Laplace's equation 
\begin{equation}
    \pd{}{r} \left( r^2 \pd{\Pih}{r} \right) + \frac{1}{\sin\theta} \pd{}{\theta} \left( \sin\theta \pd{\Pih}{\theta}  \right) + \frac{1}{\sin^2\theta} \pdd{ \Pih }{\varphi} = 0,
\end{equation} and therefore must be of the form 
\begin{equation}
    \label{eq:Pidef}
    \Pih = P(r) \Ylm(\theta, \varphi), \quad \text{ with } \quad P(r) = A_1 r^\ell + A_2 r^{- (\ell + 1)} , 
\end{equation} where $A_1$ and $A_2$ are constants of integration. As per the velocity field, suggested by~\cite{chandrasekhar} and~\cite{prosperetti1977} the problem is rephrased in terms of the perturbed vorticity $\omh = \nabla \wedge \Uh$, which is further decomposed into its poloidal and toroidal components $\omh = \bm{S} + \bm{T}$.
% \footnote{This decomposition provides a useful mathematical framework to study both the linear and weakly non-linear dynamics of spherical flows.} 
By leveraging the fact that $\nabla \wedge \nabla \wedge \omh = 0$, as provided by taking the curl of~\eqref{eq:stokeslin}, and letting $\bm{S} = \nabla \wedge \nabla \wedge \left[ S(r) \Ylm \bm{e}_r \right]$ and $\bm{T} = \nabla \wedge \left[ T(r) \Ylm \bm{e}_r \right]$, where $S(r)$ and $T(r)$ are some scalar-valued functions, it can be seen after some algebra that both satisfy the same differential equation, namely,
\begin{equation}
    D_\ell S = D_\ell T = 0, \quad D_\ell  =  \dd{^2}{r^2} - \frac{\ell (\ell + 1)}{r^2} ,
\end{equation} which are of the Euler type. The general solutions are $S(r) = A_3 r^{-\ell} + A_4 r^{\ell+1}$ and $T(r) = A_5 r^{-\ell} + A_6 r^{\ell+1}$, where $A_3$, $A_4$, $A_5$ and $A_6$ are constants of integration. The complete determination of the perturbed velocity field is then available in the form
\begin{equation}
\label{eq:u}
\Uh = T(r) \Ylm \bm{e}_r + \nabla \wedge \left[ S(r) \Ylm \bm{e}_r \right] - \nabla \Phi, \end{equation} up to a scalar gauge $\Phi$ so as to guarantee the solenoidality of $\Uh$. Such condition yields the Poisson equation
%Note that~\eqref{eq:u} is nothing but a statement of Helmholtz's theorem in which the second term is a divergence-free vector field whereas the remaining terms must satisfy the Poisson equation 
\begin{equation}
    \nabla^2 \Phi = \nabla \cdot \left[T(r) \Ylm \bm{e}_r\right],
\end{equation} which can be solved by means of separation of variables by letting $\Phi = F(r) \Ylm$, where the separation function $F(r)$ satisfies the non-homogeneous equidimensional equation
\begin{equation}
    \dd{}{r} \left(r^2 \dd{ F }{r} \right) - \ell(\ell+1) F  = \dd{}{r} \left(r^2 T\right),
\end{equation} whose solution is 
$
F(r) = A_7 r^{\ell} + A_8 r^{-(\ell+1)} + F_p, 
$ where
\begin{equation*}
    F_p = 
    - r^{\ell}      \int^r_\eta \frac{ \hat{r}^{-(\ell+3)}  }{W} \dd{}{\hat{r}}(\hat{r}^2 T) \, \mathrm{d} \hat{r} +   
      r^{-(\ell+1)} \int^r_\eta \frac{ \hat{r}^{\ell - 2}   }{W} \dd{}{\hat{r}}(\hat{r}^2 T) \, \mathrm{d} \hat{r}
\end{equation*} is the particular solution and $W = -(1+2\ell)/r^2$ is the Wronskian of the two homogeneous solutions. This, combined with~\eqref{eq:u}, finally provides the components of the perturbed velocity field $\Uh$ up to the determination of the shape functions $T$, $S$, and $F$, namely,
\begin{align}
    \urh & =  (T - F') \Ylm, \label{eq:compuh} \\
    \uth & = \frac{S}{r \sin{\theta} } \pd{\Ylm}{\varphi} - \frac{F}{r} \pd{\Ylm}{\theta}, \label{eq:comput} \\
    \uph & = -\frac{F}{r\sin{\theta} } \pd{\Ylm}{\varphi} - \frac{S}{r} \pd{\Ylm}{\theta}, \label{eq:compuphi}
\end{align} where the primes indicate differentiation with respect to the radial coordinate. The constants of integration $A_1$ and $A_2$ can be written in terms of $A_5$ and $A_6$ by simply introducing~\eqref{eq:u} into the radial component of~\eqref{eq:stokeslin} and matching powers of $r$ to find $A_1 = -(1+\ell) A_6$ and $A_2 = A_5 \ell$. The description is then completed by imposing the boundary conditions
\begin{equation}
    \label{eq:eigbcnonslip}
    \urh = \uth = \uph = 0,
\end{equation}
as required by non-slippage at the substrate wall $ r = \eta $ and 
\begin{align}
       & \urh -  \omega \zh = 0, \label{eq:eigbckinem} \\
     - & \ph + 2 \pd{\urh}{r} + \frac{\ell (\ell +1)  - 2}{\Ca  \, (\eta+1)^2} \zh  = 0, \label{eq:eigbcstress} \\
       & r \pd{}{r} \frac{\uth}{r} + \frac{1}{r}\pd{\urh}{\theta} = 0, \label{eq:eigbcrtheta} \\
       & \frac{1}{r\sin\theta} \pd{\urh}{\varphi} + r \pd{}{r} \frac{\uph}{r} = 0, \label{eq:eigbcrphi}
\end{align} 
at $r = \eta+1$, accounting for kinematic, normal- and tangential-stress equilibrium. Note that by virtue of~\eqref{eq:Pidef} and~\eqref{eq:compuh}--\eqref{eq:compuphi}, this amounts to imposing boundary conditions on the generating functions $P$, $T$, $S$, and $F$. The no-slip condition together with zero tangential stress produce $A_3 = A_4 = 0$ and, consequently, the poloidal part of the velocity perturbation vanishes, i.e., $S(r) = 0$. The problem may be then recast into the system of equations $\tens{M} \cdot \bm{A} = 0$ where $\tens{M}$ is a $4\times 4$ matrix and $\bm{A} = (A_5, \ldots, A_8)^\mathrm{T}$ is the vector of constants of integration. Finding nontrivial solutions to this system of equations needs $\operatorname{det}(\tens{M}) = 0$ thus rendering the dispersion relation in the form $\mathcal{D}(\omega, \ell, m; \eta, \Ca) = 0$ where $\omega$ emerges as a temporal eigenvalue for discrete values of $\ell$ and $|m| \leq \ell$ for given $\eta$ and $\Ca$.

\section{Functions appearing in the Stokes dispersion relation}\label{app:D}
The explicit expressions for the functions $B_i$ in Eq.~\eqref{eq:omspherical} are given by
\begin{align}
    B_1 & = \left[ 2 \eta^2 (2 \eta +3)  + 4 \eta + 1   \right]   (2 \ell+1) \eta ^{2 \ell} (\eta+1)^{2 \ell}+2 (\eta+1) \eta ^{4 \ell+3}-2 \eta  (\eta+1)^{4 \ell+3}  ,  \\
    B_2 & = 2  (\eta+1)  \left(2 \ell^2+1\right) (\ell+2) \ell \eta ^{4 \ell+3}, \\
    B_3 & = 3 (\eta+1)^4+4 (2 \eta +1)^2 \ell^6+12 (2 \eta +1)^2 \ell^5, \\
    B_4 & = \left[ 4 \eta (\eta+1) \, (2 \eta (\eta+1)  +5)  +5 \right] \ell^4 + 2 \left[4 \eta  (\eta+1)  \, (2 \eta  (\eta+1)-5)-5 \right] \ell^3, \\
    B_5 & = \ell^2 \left[ 4 \eta^2 (\eta+1) (\eta +7) + 12 \eta  + 3 \right], \\
    B_6 & =    2 \ell \eta  \left[  20  -   2\eta (\eta -6)  (\eta +2) \right] + 10  \ell   , \\
    B_7 &  =  2 \eta \,  (\ell-1) (\ell+1) \left[ 2 (\ell+2) \ell+3 \right] (\eta+1)^{4 \ell+3}.
\end{align}

\section{A simple regularization of the lubrication model}
\label{app:E}
As noted in~\cite{kang2016dynamics,kang2017marangoni}, the lubrication model~\eqref{eq:lubmodel}, and the corresponding dispersion relation~\eqref{eq:lubdisprel}, become singular in the limit $h = \eta$, which is manifested by the in-plane velocities~\eqref{eq:vtheta_lub}--\eqref{eq:vphi_lub} having a factor $h-\eta$ in their denominators. It is noted therein that this singularity is of purely mathematical origin and does not correspond to any physical mechanism. One way to remedy this is to simply invoke the fact that $\eta \gg 1$, which ultimately leads to the derivation of the thin-film model~\eqref{eq:thinfilmmodel}. The resulting dispersion relation~\eqref{eq:reynoldsDR} is then regular but is seen to perform quite poorly for order unity values of $\eta$, as discussed in the main text.

\begin{figure}
    \centering
    \includegraphics[width=270pt]{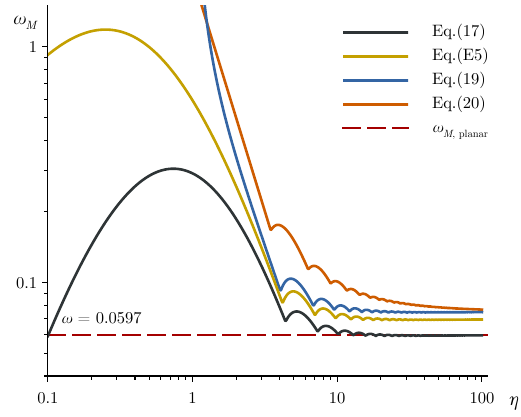}
    \caption{As in figure 3 for $Ca = 0.1$, but now including the optimal growth rate as predicted by~\eqref{eq:dispersion_relation_newlubmodel}. }
    \label{fig:figE1}
\end{figure}

A particular way of circumventing the singularity of the lubrication dispersion relation~\eqref{eq:lubdisprel} and improve on the lack of performance of~\eqref{eq:reynoldsDR} consists in incorporating information from the kinematic condition~\eqref{eq:kineminterface} into the boundary conditions for the determination of the in-plane velocities. When integrating Eqs.~\eqref{eq:lub_thetamom}--\eqref{eq:lub_phimom} with the no-slip condition, $v_\theta(\eta) = v_\varphi(\eta) = 0$, and vanishing tangential stress at $r = \eta + h$,
\begin{align}\label{eq:newstress}
    & \pd{v_\theta}{r} - \frac{v_\theta}{\eta + h} + \frac{1}{\eta+h} \frac{\partial^2 h}{\partial \theta \partial t} = 0, \\
    & \pd{v_\varphi}{r} - \frac{v_\varphi}{\eta + h} + \frac{\csc{\theta}}{\eta + h} \frac{\partial^2 h}{\partial \theta \partial t} = 0, 
\end{align} 
which is rewritten with advantage taken from the approximation $v_r \simeq \pd{h}{t}$, which is exact in linearised form~\eqref{eq:eigbckinem}, as dictated by continuity. It is important to emphasize that Eq.~\eqref{eq:newstress} is not asymptotically consistent, since the terms involving the radial velocity are smaller than those involving the polar and azimuthal velocities by a factor on the order of the flow slenderness, which is assumed to be small. The resulting lubrication model becomes
\begin{equation}
    \label{eq:newlubmodel}
    (\eta + h)^2\sin\theta \pd{h}{t} = 
    \pd{}{\theta} \left[  \sin\theta 
    \left(  G(h,\eta) \pd{p}{\theta} + H(h,\eta) \frac{\partial^2 h}{\partial \theta \partial t}   \right)    \right] + \frac{1}{\sin\theta}
    \pd{}{\varphi} \left(   G(h,\eta) \pd{p}{\theta} + H(h,\eta) \frac{\partial^2 h}{\partial \theta \partial t}  
    \right),
\end{equation} 
where 
\begin{equation}
    H(\eta,h) = \frac{3 h^2 (h+\eta)}{6 (h-\eta)}
\end{equation} 
is a new mobility function. Note that Eq.~\eqref{eq:newlubmodel} is still singular when $h \to \eta$. However, introduction of normal modes leads to a new dispersion relation given by
\begin{equation}
    \label{eq:dispersion_relation_newlubmodel}
    \omega = \frac{(2\eta+1) \, \ell (\ell + 1)  \left[  2 - \Ca \, \phi'(1) (1+\eta)^2 - \ell(\ell+1)  \right]}{3 Ca \, (1+\eta)^3 \left[  2 (\eta^2 - 1) + \ell(\ell+1) \right]},
\end{equation}
which is regular for any value of $\eta$. The performance of this dispersion relation is shown in Fig.~\ref{fig:figE1} where the optimal growth rate obtained with the regularized model is represented for $\Ca = 0.1$ as a function of $\eta$ with a yellow solid line, along with those from the main paper sections. Although the agreement is seen to improve the performance is still quite poor for $\eta\lesssim 1$, so that relatively thick films must be described with use made of the full equations of motion.

\end{appendix}

%%%%%%%%%%%%%%%%%%%%%%%%%%%%%%%%%%%%%%%%%%%%%%%%%%%%%%

\bibliography{biblio}% Produces the bibliography via BibTeX.

\end{document}